# Monte Carlo simulations for Ising spins with S>½ applied to the square and triangular lattices with antiferromagnetic interactions and comparing results using Kawasaki and Glauber dynamics

J. Poulton[1], C. Bentham[1], S.P. George[2] and G.A. Gehring[1]*

1. Department of Physics and Astronomy, University of Sheffield, Sheffield S3 7RH UK

2. CERN, Route de Meyrin 385, 1217 Meyrin Switzerland

* corresponding author g.gehring@sheffield.ac.uk

**Abstract**

This paper has a pedagogical introduction. We describe the correct method for performing Monte Carlo simulations of Ising model systems with spin greater than one half. Correct and incorrect procedures are clearly outlined and the consequences of using the incorrect procedure are shown. The difference between Kawasaki and Glauber dynamics is then outlined and both methods are applied to the antiferromagnetic square and triangular lattices for $S$ =1.

## 1. Introduction

The Monte Carlo technique is widely used to investigate the thermodynamic behaviour of large spin systems [1,2,3]. It is one of the most accurate methods of obtaining the critical behaviour of spins on frustrated lattices and for studying phase transitions. References [4,5,6,14] describe systems with $S > \frac{1}{2}$ but most previous simulations have been done for $S = \frac{1}{2}$ systems where the spin on each site has only one possible spin flip mode.

This short paper explains the flipping procedure for the general case of $S > \frac{1}{2}$ and describes the consequences of using an apparently appealing, but incorrect, procedure. Most simulations use Glauber dynamics in a single spin flip mode which explores all available phase space, however interesting effects are seen in frustrated systems using Kawasaki dynamics. In Kawasaki dynamics a pair of spins are flipped together, preserving the total magnetisation, as this is more restrictive than Glauber dynamics, it may prevent the true free energy minimum from being accessed. Results for calculations are shown first for the square lattice where there is no frustration and then for the antiferromagnetic triangular lattice which is frustrated and where there is no exact solution for either infinite spin [4] or for general spin $S > \frac{1}{2}$ [5,6].

Section two contains a description of the correct flipping procedure for systems with $S > \frac{1}{2}$ for both Glauber and Kawasaki dynamics and the effects of using an incorrect procedure. In section three both the correct and incorrect methods are applied to the antiferromagnetically coupled $S = 1$ system on a square lattice. The results of applying the correct procedure on the square lattice reproduce the known thermodynamic behaviour for $S = 1$ on a square lattice with small differences apparent from using Kawasaki dynamics. The results for the antiferromagnetic triangular lattice are presented in Section 3. As this lattice is frustrated [4,5,6] more significant differences are observed between the results obtained from Kawasaki and Glauber dynamics.

## 2. Monte Carlo Procedures and the Rate Equations

In the Metropolis algorithm for the Monte Carlo procedures sites are visited at random and the change of energy is evaluated after a spin flip [1,2]. The spin is allowed to flip if the change in energy is negative or its value is less than a random number multiplied by $k_B T$ [1,2]. This process is unique for $S = \frac{1}{2}$ but there is some ambiguity for larger spins that is addressed here.

Consider the state $|S, m\rangle$. In the general case there are two modes of change available, in which there is one unit of spin change, moving the state to $|S, m \pm 1\rangle$. However in the special case that $m = \pm S$ then one of these modes is not allowed. Here there are two available strategies; **A** the code automatically only allows certain flipping modes mode and then the normal Monte Carlo procedure is applied or **B** the direction is randomly selected and automatically fails if the move is to a state $|m'| > S$.

The correct procedure is strategy **B** because the principle of detailed balance requires that the probability of entering a state must be equal to the probability of leaving the state. It can be easily seen that procedure **B** satisfies this using the rate equations for Kawasaki and Glauber dynamics in the limits of infinite temperature. The consequences of using the incorrect procedure **A** are incorrect values of spin probabilities at high temperatures, and hence incorrect values of the entropy.

The effects of these two different procedures on the spin probabilities are seen using rate equations. These are derived below for Glauber and Kawasaki dynamics in the limit of infinite temperatures.

We consider a single spin flip for Glauber dynamics, $\Delta m = \pm 1$, for $S = 1$ and then generalize the discussion for general spin. Let $p_1$, $p_0$ and $p_{-1}$ be the probabilities finding a site with $m = \pm 1, 0$ the decay rates are given as $\alpha$ for the states $|\pm 1\rangle$ and for $\beta$ for state $|0\rangle$. The rate equations are given below in the limit of infinite temperature allowing for the branching ratio for the decay of a state $|0\rangle$.

$$\frac{dp_1}{dt} = -\alpha p_1 + \frac{\beta}{2} p_0$$

$$\frac{dp_0}{dt} = -\beta p_0 + \alpha(p_1 + p_{-1})$$

$$\frac{dp_{-1}}{dt} = -\alpha p_{-1} + \frac{\beta}{2} p_0$$

The time derivatives vanish in equilibrium leading to

$$p_1 = p_{-1} \text{ and } \alpha p_1 = \frac{\beta}{2} p_0.$$

When using, the incorrect, procedure **A** the Monte Carlo procedure has set $\alpha = \beta$ and hence $p_1 = p_{-1} = \frac{p_0}{2}$ so upon normalizing we obtain

$$p_1 = p_{-1} = \frac{1}{4} \text{ and } p_0 = \frac{1}{2}.$$

When using procedure **B** the principle of detailed balance requires that at infinite temperature $p_1 = p_{-1} = p_0$ and thus

$$p_1 = p_{-1} = p_0 = \frac{1}{3}.$$

This gives $\alpha = \frac{\beta}{2}$. Hence procedure **B** leads to an ergodic procedure and is a correct procedure whereas procedure **A** is incorrect.

It is straightforward to generalize this to the case of general spin $S > 1$ to find that for procedure **A** $p_s = p_{-s} = \frac{p_m}{2}$, $m \neq \pm S$ leading to the incorrect result;

$$p_{\pm s} = \frac{1}{4s}, p_m = \frac{1}{2s}, m \neq \pm S$$

whereas for procedure B we find the correct result

$$p_m = \frac{1}{2s+1}$$

We note that an alternative strategy is to allow each state $|S, m\rangle$ the possibility of making a transition between all the other $m$ states where the particular transition is chosen at random and where all outcomes have different probabilities. This leads to the same result as procedure **B** and is also correct.

We now consider the case of Kawasaki dynamics in which a pair of neighbouring spins makes a transition from $|m_1, m_2\rangle$ to $|m_1 \pm 1, m_2 \mp 1\rangle$. This is equivalent to allowing a small Heisenberg term to drive the dynamics of the Ising spin system. In the Monte Carlo simulation there is an extra step in which one of the possible end states is selected at random.

As before we consider the rate equations in the limit of infinite temperature for nearest neighbour pair probabilities $P_{m_1,m_2}$ and using the same arguments as above we see that the pair probabilities are

$$\alpha P_{1,-1} = \frac{\beta}{2} P_{0,0}.$$

When using procedure **A** the Monte Carlo procedure has set $\alpha = \beta$ so this leads to $P_{1,-1} = \frac{1}{2} P_{0,0}$ and hence the single site probabilities, $p$, are related by $p_1^2 = p_{-1}^2 = \frac{p_0^2}{2}$ and thus $p_1 = p_{-1} = \frac{p_0}{\sqrt{2}}$. Normalization gives

$$p_1 = p_{-1} = \frac{1}{2+\sqrt{2}} = 0.293 \text{ and } p_{-1} = \frac{1}{1+\sqrt{2}} = 0.414.$$

Whereas applying detailed balance $P_{1,-1} = P_{0,0}$ as $t \to \infty$ we find that using procedure B ($\alpha = \frac{\beta}{2}$) gives the single site probabilities as above for Glauber dynamics.

Again procedure **A** can be generalized to give the results for general spin:

$$p_s = p_{-s} = \frac{1}{\sqrt{2}(2S - 1 + \sqrt{2})}$$

$$p_m = \frac{1}{2S - 1 + \sqrt{2}}$$

Hence we have shown the error introduced by using strategy **A** affects the spin probabilities at high temperatures and hence the entropy. The entropy is lower than expected due to the favouring of the middle spins during the process. In the next sections the detailed results are shown for the case of $S = 1$ for the antiferromagnetic square and triangular lattices.

### 3. Results for antiferromagnetically coupled spins $S = 1$ on a square lattice

We test our procedure on the well understood antiferromagnetic $S = 1$ square lattice that has a fully ordered ground state containing two sublattices. We give results for probability, susceptibility and heat capacity and discuss the process of calculating entropy and correcting for finite size and temperature effects. The Hamiltonian is given by

$$H = J \sum_{<i,j>} m_i m_j, \ m_{i,j} = 0, \pm 1$$

For the Ising model the thermodynamics of the square antiferromagnet is exactly equivalent to that of the square ferromagnet. Most real magnets are described well using Glauber dynamics however ordering of side chains on a liquid crystal square lattice leads to an antiferromagnet with *S*= 2 which necessarily requires Kawasaki dynamics because a flip of one chain changes the order on two neighbouring sites [15, 16].

Monte Carlo calculations have been performed using procedures **A** and **B** (as above) in order to study the effects of using Glauber and Kawasaki dynamics for this model. The simulations were done on a 12 by 12 lattice of spins with periodic boundary conditions. The simulations started with the system in the known ground state and were performed as a heating run. The spins were thermalized using 10000n (where n is the number of lattice sites) iterations per temperature step with a further 1000 data collection iterations. This is necessary because if there is any remaining correlation between temperature steps then this will lower the entropy and give false results.

The number of thermalisation iterations required to make each temperature step completely independent is dependent on a correlation function which is given by $\chi = e^{-\frac{t}{\tau}}$ where it is found that $2\tau$ iterations is suitable for full thermalisation [2].

Procedure **A** will give a result for entropy which is significantly too low compared to procedure **B** due to the favouring of the $m = 0$ state for a spin one lattice or the $m \neq \pm S$ state for general spin greater than one.

When calculating entropy, we have to correct for finite size effects and finite temperature effects. The finite size effects can easily be corrected for by running the model for a number of lattice sizes and then plotting the graph of entropy against $\frac{1}{L^2}$ where L is the number of sites per side (or the length). This line can then be extrapolated to $L = \infty$ to give the infinite lattice result.[14]

The entropy approaches the high temperature limit with a correction term that varies as $\frac{1}{T^2}$. Hence the high temperature limit may be found by plotting the entropy found at temperature T against $\frac{1}{T^2}$. The following figure shows the results of using a square lattice using Glauber dynamics that has not been corrected. It can be seen that the entropy at infinite temperature is lower than the expected value of $ln3$.

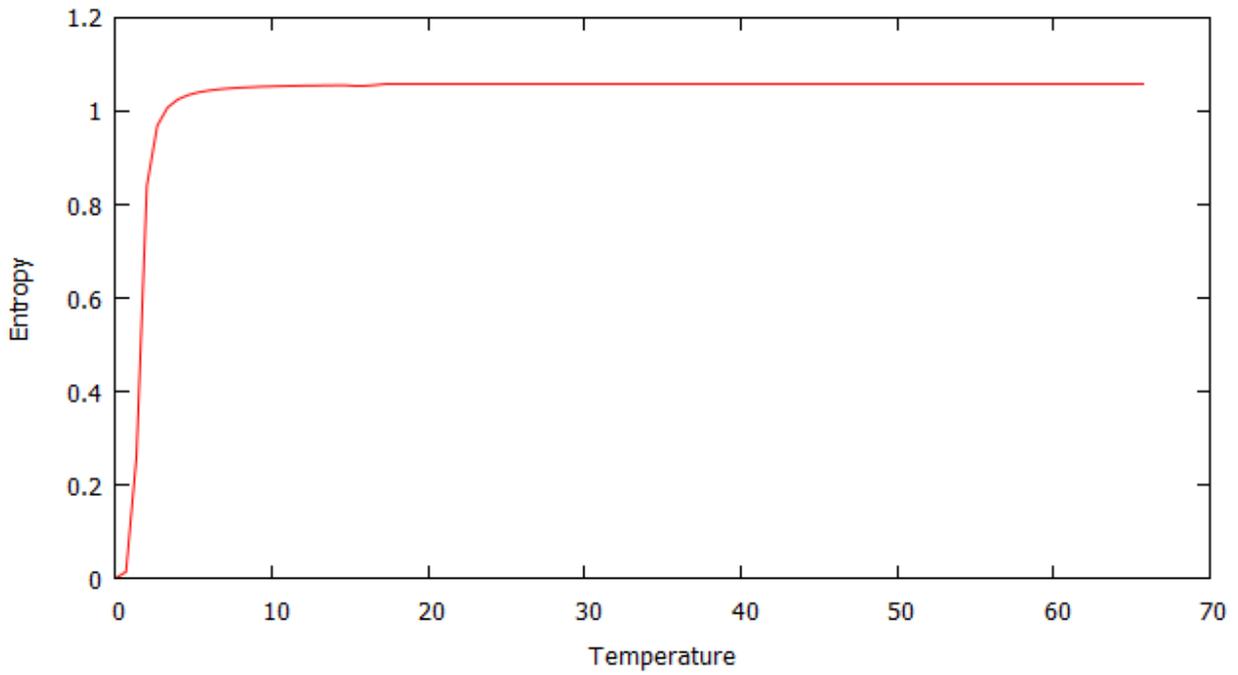

Figure 1 The entropy of the Glauber model using procedure B. It can be seen that even at high temperature the entropy has not reached ln3 due to finite size effects (this is a 12x12 lattice).

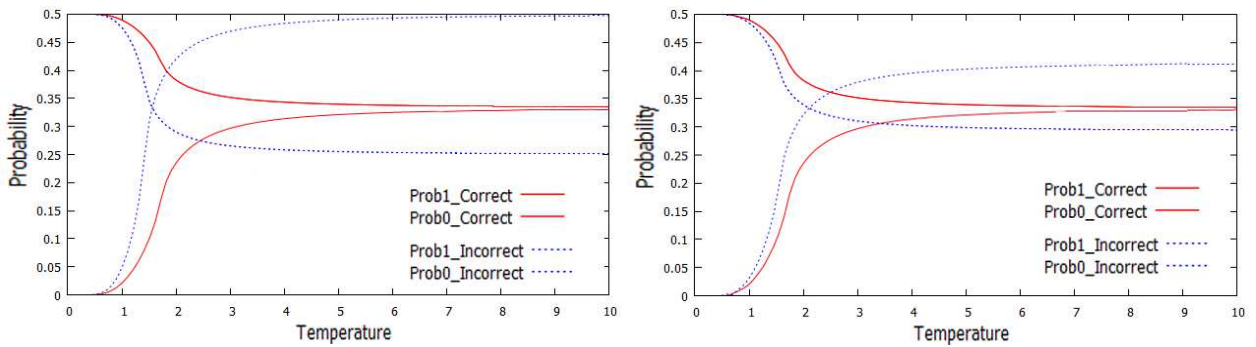

Figure 2 The temperature dependence of probability for a S=1 antiferromagnet on a square lattice using i) Glauber Dynamics and ii) Kawasaki Dynamics. In both cases the correct procedure B is shown with solid lines and the incorrect procedure A with dashed lines.

In Figure 2 we show the evolution of the probabilities $(p_{\pm 1}, p_0)$ as a function of temperature. At $T = 0$ we have $p_{\pm 1} = \frac{1}{2}$ and $p_0 = 0$ as expected for full order.

In the high temperature limit using procedure **B** we find that $p_1 = p_{-1} = p_0 = \frac{1}{3}$ as expected, for both Kawasaki and Glauber dynamics. However for procedure **A** we find the probabilities tend to 0.25 and 0.5 for Glauber dynamics and 0.293 and 0.414 for Kawasaki dynamics in the limit of high temperatures. This corresponds to the probabilities as calculated in the previous section. These results naturally lead to incorrect values of the change in entropy from low to high temperatures.

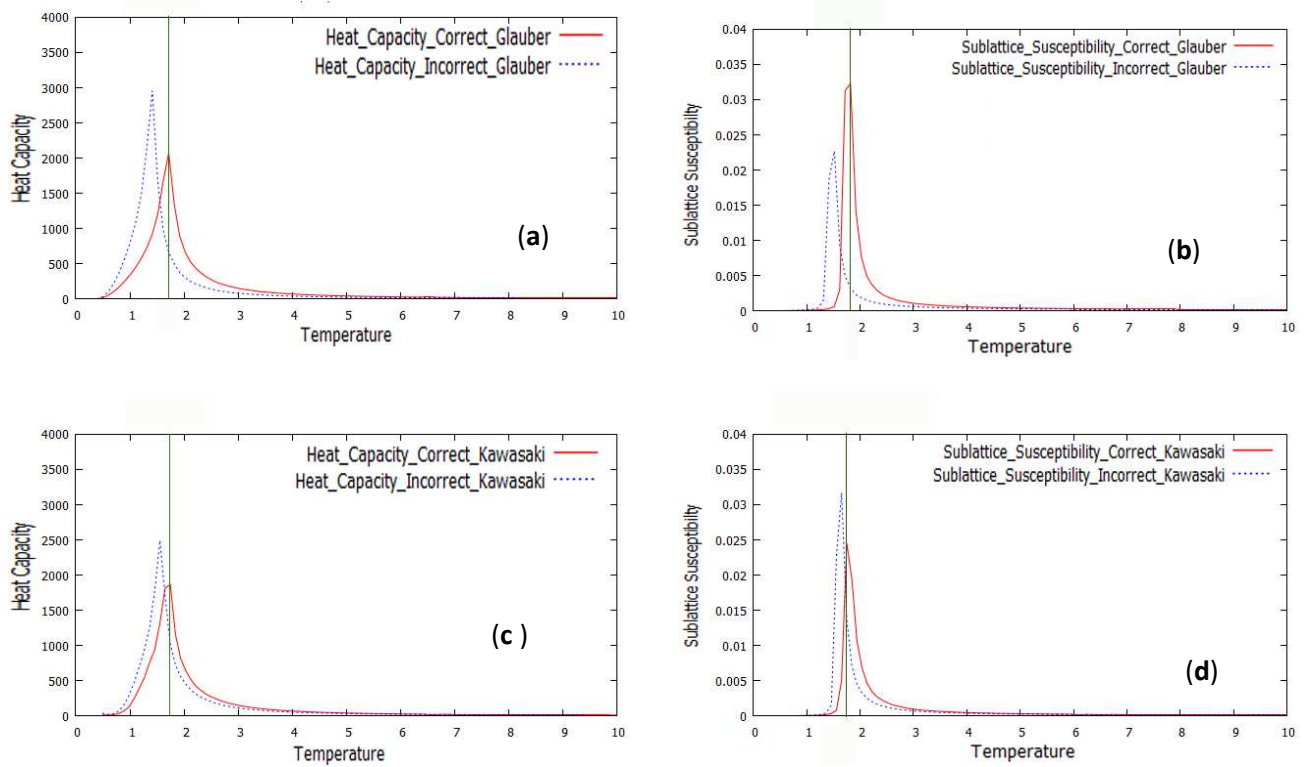

Figure 3. The results for Glauber dynamics for (a) the heat capacity and (b) the susceptibility and for the (c) heat capacity and (d) sublattice susceptibility using Kawasaki dynamics. The critical temperatures, as indicated by the maximum values of the heat capacity and susceptibility are indicated by dotted lines.

The transition temperature for the $S = 1$ Ising model on a square lattice has been evaluated to be $k_B T_c \cong 1.6936$ using a low temperature series expansion and a transfer matrix calculation [7,8]. In Figures 2 and 3 we show the results of our simulations for the specific heat and sublattice susceptibilities for Glauber and Kawasaki dynamics in order to check the dependence of physical properties near the transition on the type of dynamics used and the effects of using the incorrect procedure **A**. The values of the critical temperature are best estimated from the heat capacity and susceptibility graphs.

It is well known that different critical dynamics is expected for Glauber and Kawasaki because the Kawasaki method conserves the total z component of spin [9]. We see here that there are also small changes in the static critical properties. The values of the critical temperature obtained in these simulations are $1.71 \pm 0.02$ for Glauber and $1.74 \pm 0.02$ for Kawasaki with the Glauber result within one standard deviation of the accepted result and the Kawasaki just outside the second.

### 3. Results for antiferromagnetically coupled spins S = 1 on a triangular lattice

The antiferromagnetic lattice with $S = \frac{1}{2}$ has been solved exactly [10] and has no long range order. For large spin $S > 3$ there is there is an entropy driven transition to an ordered state with three sublattices. These sublattices are $m_1 = S$, $m_2 = -S$ and $\langle m_3 \rangle = 0$. The spins on the third sublattice have three neighbours with m $= +S$ and three with m $= -S$ and hence can take all available states; $-S \leq m \leq S$.

A spin on any given sublattice has 3 neighbours on each of the two other sublattices, e.g. a spin on sublattice 1 has 3 neighbours on each of sublattices 2 and three. However if three spins on the mixed lattice (sublattice 3) surrounding a single site on sublattice 1 or 2 take the same value as this central spin, they will destabilize this spin, creating a defect[5]. This has the effect of triggering an interchange between the third sublattice and one of the other two. A recent work for $S = 1$ [13] showed that the effect of a single ion anisotropy, that disfavours the states *m*= ± 1 could lead to full order even for *S* =1.

The difference between the probabilities of obtaining the states *m = ±1* and *m=0* for Glauber and Kawasaki dynamics are negligible; however the sublattice susceptibility shows a large dependence on the type of dynamics. At low temperatures many transitions are frozen out when using Kawasaki dynamics [4]. This is especially severe for Kawasaki dynamics for a frustrated model because since two neighbours must flip together this means that the flips must occur on two different sublattices [15, 16]. Many such flips require an energy input and so are frozen out at low temperatures. Thus it becomes harder to reach a true equilibrium with Kawasaki dynamics than with Glauber dynamics.

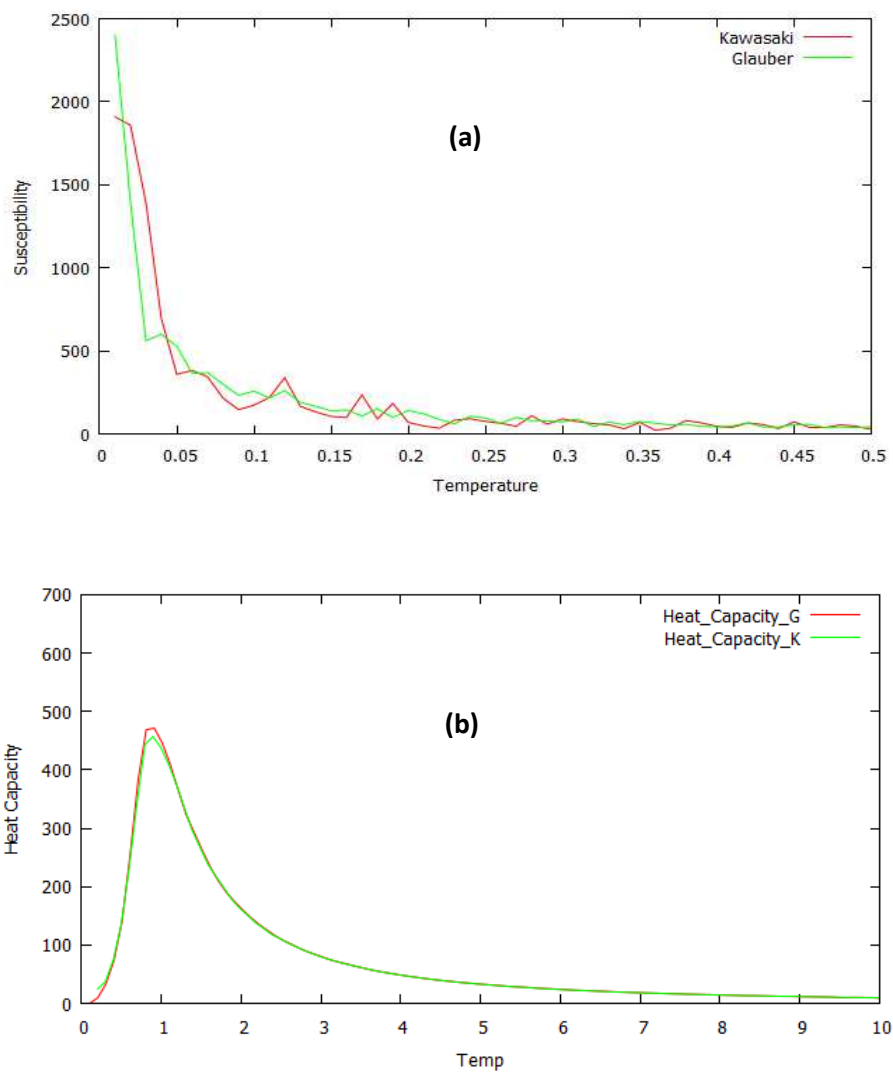

**Figure 4 Comparison of (a) the low temperature susceptibilities and (b) the specific heats for Glauber and Kawasaki dynamics.**

Figure 4 (a) shows much a sharper peak for the susceptibility using Kawasaki dynamics due to the much longer correlation rates due to slow decay, the susceptibility is dominated by correlations at long range. At very low temperatures the simulation using Kawasaki dynamics develops such a long relaxation time that it can no longer reach lower energy levels whereas the Glauber susceptibility becomes much higher. Neither Glauber nor Kawasaki dynamics lead to long range order. Instead they form a three lattice state similar to that expected for the ground state for large spin (one m=S, one m=-S and one mixed) however the three sublattices constantly interchange due to defect proliferation. This can continue down to zero using Glauber dynamics whereas in the Kawasaki model, this exchange is frozen out, and is forced into a state which is not the true ground state. By contract the specific heat which depends on nearest neighbour correlations only shows almost no difference between Kawasaki and Glauber dynamics.

**Conclusions**

The differences between correct and incorrect procedures are strongly apparent in the results of the simulations. The state probabilities tend to incorrect values at high temperatures and hence incorrect values for the entropy.

However the differences between using Glauber and Kawasaki correctly are more subtle. For the square lattice there is a small difference in the calculated transition temperatures and only the sublattice susceptibility show a dramatic difference for the triangular lattice.

**Acknowledgements**

We would like to thank Professor Henley for helpful correspondence and the EPSRC for a summer bursary for JP and a research studentship for CB. SPG has been supported by the Marie Curie Initial Training Network Fellowship of the European Community's Seventh Framework Programme under Grant Agreement PITN-GA-4 2011-289198-ARDENT.